\documentclass[a4paper,11pt]{article}
\pdfoutput=1 % if your are submitting a pdflatex (i.e. if you have
\usepackage{jheppub} % for details on the use of the package, please
\usepackage{comment}

\numberwithin{equation}{section}

\gdef\ffrac#1#2{\textstyle\frac{#1}{#2}\displaystyle}
\gdef\be{\begin{equation}}
\gdef\ee{\end{equation}}
\gdef\e{\epsilon}

\gdef\p{\partial}
\gdef\ve{\varepsilon}

\gdef\zb{{\bar z}}
\gdef\d{{\delta}}

\title{\boldmath $T\overline T$ deformation of correlation functions}

\author[a,b]{John Cardy}

\affiliation[a]{Department of Physics, University of California, Berkeley CA 94720, USA}
\affiliation[b]{All Souls College, Oxford OX1 4AL, UK}

\emailAdd{cardy@berkeley.edu}

\abstract
{We study the evolution of  correlation functions  of local fields in a two-dimensional quantum field theory under the $\lambda T\overline T$ deformation, suitably regularized. We show that this may be viewed in terms of the evolution 
of each field, with a Dirac-like string being attached at each infinitesimal step.  The deformation then acts as a derivation on the whole operator algebra, satisfying the Leibniz rule.
We derive an explicit equation which allows for the analysis of UV divergences,
which may be absorbed into a non-local field renormalization to give correlation functions 
which are UV finite to all orders, satisfying a (deformed) operator product expansion and a Callan-Symanzik equation. We solve this in the case of a deformed CFT, showing that the Fourier-transformed renormalized two-point functions behave as $k^{2\Delta+2\lambda k^2}$,
 where $\Delta$ is their IR conformal dimension. We discuss in detail deformed Noether currents, including the energy-momentum tensor, and show that, although they also become non-local, when suitably improved they remain finite, conserved and satisfy the expected Ward identities. 
Finally, we  discuss how the equivalence of the  $T\overline T$ deformation to a state-dependent coordinate transformation emerges in this picture.}

\begin{document} 
\maketitle
%\flushbottom

\section{Introduction}\label{sec:intro}

In recent years the deformation of  a 2d local quantum field theory by a term in the action proportional to the determinant $\det T$ of the stress tensor, commonly referred to as $T\overline T$ \cite{Zam1}, has been of interest for a number of reasons. It gives an example of a UV completion of such a theory which is not itself a local QFT, although it retains several desirable features of such (as well as some undesirable ones). It is equivalent to coupling the theory to a form of 2d gravity \cite{Dub}. Within the framework of AdS/CFT duality, there is strong evidence that it corresponds to moving a finite distance into the AdS$_3$ bulk from the boundary CFT \cite{Verl,Kraus}. 

In addition, some properties of the deformed theory are exactly calculable given the data of the undeformed theory, whether it is integrable or not. 
Despite the fact that the deformation parameter $\lambda$ has the dimensions of (length)$^2$ and the deformation is  apparently non-renormalizable, many quantities are UV finite, including the partition function on a torus \cite{JC1,Datta} and the related finite-size energy spectrum \cite{Zam1}, the thermodynamics, the spectrum of asymptotic states in a massive theory, and their $S$-matrix which acquires only CDD phase factors \cite{Dub2,Zam2,Tat1}. 

This solvability of the deformation may be explained in various equivalent ways, including but not limited to:
\begin{enumerate}
\item factorization, that is the factors in $\det T$, which is quadratic in the components of $T$, are independent of their separation, up to derivatives of local fields \cite{Zam1}; 
\item the deformation is equivalent to coupling the theory to Jackiw-Teitelboim \cite{Jac,Tei} gravity \cite{DubJT};
\item the deformation is equivalent to coupling the theory to a random flat metric, whose action is a total derivative \cite{JC1};
\item $\det T$ itself is a total derivative if the conserved current $T$ is expressed as the curl of a (semi-local) field \cite{JC2}.  In this paper we shall use this method, and also describe a fourth based on Green functions (Sec.~\ref{sec:GF}).
\end{enumerate}

The fact that, for example, the spectrum becomes independent of the finite size $R$ for $|\lambda|\gg R^2$ suggests that on this length scale the deformed theory is non-local. It is therefore an interesting question to understand the fate of local fields and their correlation functions under the deformation. Unlike the above global quantities these are strongly UV divergent and require a short-distance cut off $|\ve|\sim\Lambda^{-1}$ to make them finite. The question is then whether there exists a sensible non-trivial renormalized theory in which the cut off may be removed. Usually for a non-renormalizable deformation this is not the case, as UV divergences proliferate uncontrollably at higher orders in perturbation theory. However, we shall argue that, because of the above special properties of the $\det T$ deformation, the important divergences are controllable, and, moreover, have a nested form which allows for a renormalization procedure, albeit unconventional, .

Our analysis is based on the evolution equation for a general correlation function in ${\mathbb R}^2$ of fields $\{\Phi_n(x_n)\}$
 in the deformed theory \cite{Zam2}:
\be
\p_\lambda\langle\prod_n\Phi_n(x_n)\rangle_\lambda=\int\langle\det T^\lambda(x)\prod_n\Phi_n(x_n)\rangle^c_\lambda d^2x\,,
\ee
where $\det T$ is regularized by point-splitting. Our first result is that this may be cast in terms of  the evolution of each field:
\be\label{1.2}
\p_\lambda\langle\prod_n\Phi_n(x_n)\rangle_\lambda
=\sum_n\langle \p_\lambda\Phi^\lambda_n(x_n)  \prod_{m\not=n}\Phi^\lambda_m(x_m)\rangle_\lambda\,,
\ee
that is, the deformation $\p_\lambda$ satisfies the Leibniz rule  acting on a product of local fields, and is therefore a derivation on the associative algebra of such fields. This property is a consequence of the partial solvability of the $T\overline T$ deformation, and would not hold for a general deformation.
Moreover we have the explicit formula
\be\label{1.3}
\p_\lambda\Phi^\lambda(x)
=2\pi\e^{ab}\e^{ij}\int_x^X T^\lambda_{ai}(x'+\ve)dx'_j\,
\p_{x^b}\Phi^\lambda(x)\,,
\ee
where $\ve$ is the point-splitting regulator, and the integral is along any path from $x$ to an arbitrarily chosen point $X$ which avoids all the other arguments.
The full result (\ref{1.2}) is then independent of the paths and of $X$.

The fact that the insertion of $\det T^\lambda$ integrated over ${\mathbb R}^2$ may be reduced to an integration of a single insertion of $T^\lambda$ along a one-dimensional curve is a result of the partial solvability of the deformation. (\ref{1.3}) is non-trivial because it involves the updated stress tensor $T^\lambda$ of the deformed theory, which itself obeys an equation of the same form, 
with an additional term arising from the explicit change in the action, which is necessary to ensure continued conservation of $T^\lambda$.

Eq.~(\ref{1.3}) allows us to analyze the UV divergences order by order in $\lambda$, since new ones arise only from the $\ve\to0$ limit, and thus their form may be deduced from knowledge of the OPE between $T^\lambda(x')$ and $\Phi^\lambda(x)$. Although in principle this may contain arbitrarily high negative powers of $|x'-x|$, leading to potential power law divergences, as for a conventional non-renormalizable deformation, we argue that these are in fact absent in this regularization scheme, a result which is confirmed by a first order calculation about a CFT.

The $O((x'-x)^{-1})$ terms in the OPE, however, lead to more interesting logarithmic divergences. They are fixed by the Ward identity 
and rotational symmetry, and are therefore universal. 
For a generic 
field we then find that 
\be\label{1.5}
\p_\lambda\Phi^\lambda(x)=(\log\Lambda)\nabla^2_x\Phi^\lambda(x)+\mbox{less divergent terms.}
\ee
 The appearance of such logarithmic divergences in first order perturbation theory was noticed in \cite{Kraus}, but here we find that they occur to all orders, and moreover, the coefficient is universal and independent of $\lambda$ since it is fixed by the Ward identity. If we now ignore the remainder and solve (\ref{1.5}) in Fourier space
\be\label{1.6}
 \Phi^\lambda(k)=e^{-\lambda\log(\Lambda/\mu) k^2}\,\Phi^0(k)
 \ee
 (where $\mu$ is some arbitrary normalization scale) this resums all the leading terms of the form $\lambda^N(\log\Lambda)^{N'}$ with $N'=N$: the remainder all have $N'<N$. Thus there is a non-trivial scaling limit where
 $\Lambda\to\infty$ with $\lambda\log\Lambda$ fixed in which all the other terms vanish and the deformed correlation functions are simply those of the undeformed theory convoluted with the Fourier transform of the heat kernel in (\ref{1.6}). 
 Thus, in this limit, and for $\lambda>0$, the deformation is equivalent to the arguments $x_n$ of the fields executing independent Brownian motions with a diffusivity $O(\log\Lambda)$.\footnote{This is different from the diffusive motion in moduli space of the partition function discussed in \cite{JC1}, which has diffusivity $O(1)$.} 
 
To go beyond this weak-coupling limit, we may instead try to define renormalized fields by inverting (\ref{1.6});
\be
\widehat\Phi^\lambda(k)\equiv e^{\lambda\log(\Lambda/\mu) k^2}\,\Phi^\lambda(k)\,.
\ee
Using (\ref{1.5}) we may then show that correlators of $\widehat\Phi^\lambda$ are indeed finite as the regulator $\ve\sim\Lambda^{-1}$ is removed. Moreover they satisfy a deformed version of the OPE, and both correlators and OPE coefficients satisfy a Callan-Symanzik-type RG equation. The solutions, however, are peculiar: for example the two-point functions of a scalar field in a deformed CFT behave in $k$-space like
\be\label{2ptk}
\widehat C(k)\propto e^{2\lambda(\log(k^2/\mu^2)k^2}
\ee\,.
Taking the Fourier transform of this is difficult, but in Sec.~\ref{sec:RG} we argue that it gives an asymptotic expansion in powers of $\lambda\log |x|/x^2$ for large $x$, while for $|\lambda|\gg|x|^2$ the correlation function behaves like
$e^{-x^2/4\lambda\log(4\lambda^2\mu^2/x^2)}$ for $\lambda>0$ (the case with Hagedorn behavior), while it oscillates on the scale $\mu^{-1}$ for $\lambda<0$.

 However an assumption in the above analysis, indeed in all the literature on $T\overline T$,  is that the deformed stress tensor itself $T^{(\lambda)}$ continues to have its usual properties, in the sense that it is finite (up to possible derivative terms which may be removed by improvement), remains conserved, and satisfies the correct Ward identities. We argue that this is indeed the case for any Noether current $J_c$ corresponding to a symmetry of the deformed action. This is despite the fact that such a field satisfies the same non-local evolution equation (\ref{1.3}). However in general there is an additional contribution to the current coming from the explicit deformation of the action, which ensures its continued conservation. It turns out that this extra term is such as to modify (\ref{1.3}) to
 \be
 \p_\lambda J^c(x)=2\pi\e^{cb}\e^{ij}\p_{x^b}\int_x^Xdx'_jT^\lambda_{ai}(x'+\ve)J_a^\lambda(x)\,,
 \ee
 so that, although this is UV divergent as $\ve\to0$, these are in total derivatives and moreover do not affect the divergence of the current, so it continues to satisfy its Ward identity. The same is true  for the deformed stress tensor.
 
 However, the derivative $\p_l$ in Eq.~(\ref{1.3}) also suggests another interpretation. Instead of deforming the field we deform its argument: $\p_\lambda\Phi^\lambda_n(x_n)=\Phi^\lambda_n(\p_\lambda x_n)$ where
 \be
 \p_\lambda x^l_n=2\pi \int_{x_n+\ve}^X\e^{kl}\e^{ij}T^\lambda_{kj}(x')\,dx'_i\,.
 \ee
 This is a strange looking equation, as it appears to imply a field-valued coordinate transformation, but it makes at least formal sense in correlation functions. Moreover if we quantize the theory on $x^0=$ constant, and run the integration along this axis, with $X=\infty$, this becomes
 \be
 \p_\lambda x^1_n=-2\pi \int_{x^1_n+\ve}^\infty T^\lambda_{00}(x')\,dx'_1\,,\qquad
\p_\lambda x^0_n=2\pi \int_{x+\ve}^\infty T^\lambda_{10}(x')\,dx'_1\,,
 \ee
 so the spatial coordinate $x_n^1$ gets shifted by an amount proportional to the integrated energy density to the right of the point $x_n$, while the time coordinate $x_n^1$ gets shifted by an amount proportional to the integrated momentum density to its right. In the Hilbert space formulation, these may then be viewed as a state-dependent diffeomorphism, an interpretation already pointed out for classical theories in \cite{Conti1,Conti2}. It is also consistent with the form of the CDD factors
 $e^{-i\lambda\sum_{a,b}\e^{ab}p_a^0p_b^1}$ \cite{Dub}. 
 
 So far, the $T\overline T$ deformation of correlation functions has received relatively little attention. Kraus, Liu and Marolf \cite{Kraus} computed correlators of the stress tensor to lowest nontrivial order about a CFT, and also   the 2-point functions of generic fields to first order. Their motivation was a comparison with the holographic interpretation. Aharony and Vaknin \cite{Aharony} discussed a different limit from the present paper, in which $\lambda \to0$, $c\to\infty$, with $\lambda c$ fixed.

 The outline of this paper is as follows. In Sec.~\ref{sec:JJ} we consider the example of a deformation $J^1\wedge J^2$, where $J^1$ and $J^2$ are a pair of commuting vector symmetry currents. This is also a total derivative of a semi-local field, and so has much in common with the $\det T$ deformation but is simpler in some respects, notably in that it is a marginal rather than a UV relevant deformation. We first explore its effect on correlators in first order perturbation theory, then more generally, using the OPE. It turns out that the deformation induces logarithmic correlations between fields which carry both non-zero charge and vorticity. In fact this deformation has much in common with the $\theta$-term considered some time ago in a dimensionally reduced version of $F\widetilde F$ in 4d gauge theory \cite{EC1,EC2,BG}. In 
Sec.~\ref{sec:TT} we then repeat the exercise for the $\det T$ deformation, derive the main results (\ref{1.2}, \ref{1.3}) and extend these to the deformation of conserved currents. We then (Sec.~\ref{sec:div}) use these to analyze the UV divergences to all orders, use these to resum leading logs to find 
the diffusive scaling limit, and then show how to define renormalized fields whose correlation functions are finite to all orders. This leads to the RG analysis and the solution (\ref{2ptk}) for the 2-point function. We also discuss the deformed OPE satisfied by these renormalized fields. 
In Sec.~\ref{sec:GF} we describe an alternative method of factorizing the $\det T$ deformation, which reproduces both our Eq.~(\ref{1.3}) and also Zamolodchikov's equation \cite{Zam1} for the deformation on the cylinder, and which should be useful for other 2d manifolds. Finally in Sec.~\ref{sec:diffeo} we show how the interpretation of the deformation as a field- (or state-) dependent coordinate transformation arises from the perspective of this paper, and end with some conclusions and open problems.

\section{$J^1\wedge J^2$ deformation}\label{sec:JJ}
Before discussing the $T\overline T$ deformation it is useful to consider this simpler deformation as much of the analysis is similar.
Consider a 2d euclidean quantum field theory in flat space with two commuting conserved vector currents $J^a_i$ $(a=1,2)$, which are conserved apart from possible localized sources corresponding to operator insertions, where the charge $\p^iJ^a_i\not=0$, and point vortices, around which the circulation $\oint J^a_idx^i\not=0$. Note that although $\p^iJ^a_i=0$ except at the sources, the bulk vorticity $\e^{ij}\p_iJ^a_j$ does not vanish in general, since this would imply that the complex components $J_z$ ($J_\zb$) are (anti-)holomorphic as in a CFT. 
The action is deformed infinitesimally by a term
\be
-\delta\lambda\,\e_{ab}\int J^a(x)\wedge J^b(x)\,d^2x
=-\delta\lambda\,\e_{ab}\e^{ij}\lim_{\ve\to 0}\int J_i^a(x)\wedge J_j^b(x+\ve)\,d^2x\,,
\ee
where we have defined the product by point-splitting, in anticipation of possible divergences in correlators as $\ve\to0$.
In principle we should average over directions of $\ve$ in order to maintain rotational invariance:
\be\label{symm}
\lim_{|\ve|\to 0}\int J_i^a(x)\wedge J_j^b(x+\ve)d\ve/|\ve|
\ee
although in practice this is unnecessary (except when showing that symmetry of $T_{ai}$ is preserved by the deformation in Sec.~\ref{sec:Tdef}).  
As for the $T\overline T$ deformation, the currents, when expressed in terms of the fields of the undeformed theory, might depend on the deformation parameter $\lambda$, but it is important that they continue to be conserved. If they are Noether currents of some symmetry, this should therefore be respected by the deformation. An example would be $U(1)\times U(1)$. 
In the absence of sources and point vortices we may write, locally\be
J^a_i=\e_{ik}\p^k\chi^a\,,
\ee
where the $\chi^a$ ($a=1,2$) are semi-local scalar fields, sometimes referred  to as prime forms.\footnote{For a conserved symmetric tensor this idea goes back to Airy in 1863 \cite{Airy}. See 
\cite{Pom}.}
 In terms of these
\begin{eqnarray}
\e_{ab}J^a(x)\wedge J^b(x+\ve)&=&\e_{ab}\e^{ij}\e_{jk}J^a_i(x)\p^k\chi^b(x+\ve)\\&=&\e_{ab}J^a_i(x)\p^i\chi^b(x+\ve)
=\e_{ab}\p_x^i[J^a_i(x)\chi^b(x+\ve)]\,.
\end{eqnarray}
The main point is that this is a total derivative and, in the absence of sources, integrates to either a boundary term, or, for a closed manifold, is non-zero only when there is non-trivial homotopy allowing winding for the fields $\chi^b$. As discussed in \cite{JC2} this gives a non-zero contribution to the deformation of the torus partition function
\be
\p_\lambda\log Z=-\e_{ab}\e^{ij}\langle Q_i^aQ_j^b\rangle\,,
\ee
where $Q^a_i$ is the charge corresponding to $J^a$ circulating around the cycle $i$. 

However in this paper we consider mainly the infinite euclidean plane, where the $\lambda$-dependence of the partition function is trivial but the correlation functions of local fields with non-zero charge and vorticity are not. This is because the fields $\chi^{a,b}$ are singular at the sources of the currents, and also non-single valued due to their vorticity. In fact,
close to each singularity (chosen for convenience to lie at the origin) we have
\be\label{2.7}
J^a_i\sim(1/2\pi)(q^ax_i/x^2+\tilde q^a\e_{ij}x_j/x^2)+\cdots\,,
\ee
where $(q^a,\tilde q^a)$ are the charges and vorticity respectively. (\ref{2.7}) may also be viewed as the leading terms in the OPEs of the currents with local fields which insert the sources, and the omitted terms are less singular.
In complex coordinates,  it reads
\be
J^a_z\sim(1/4\pi)(q^a+i\tilde q^a)/z+\cdots\,,\quad J^a_\zb\sim(1/4\pi)(q^a-i\tilde q^a)/\zb+\cdots\,.
\ee

\subsection{First order deformation about a CFT}
In order to see the structure of the integral $\int \langle J^1\wedge J^2\rangle d^2x$, it is useful first to examine the first order in perturbation theory in $\lambda$ about a CFT. In complex coordinates we have 
\be
\e_{ab}\int \langle J^a(x+\ve)\wedge J^b(x)\rangle\,d^2x
=2i\e_{ab}\int\langle J^a_z(z+\ve)\rangle \langle J^b_\zb(\zb)\rangle d^2z\,,
\ee
where, by the Ward identity,
\begin{eqnarray}
\langle J^a_z(z)\rangle&=&(1/4\pi)\sum_n\frac{q_n^a+i\tilde q_n^a}{z-z_n}\,,\\
\langle J^a_\zb(\zb)\rangle&=&(1/4\pi)\sum_n\frac{q_n^a-i\tilde q_n^a}{\zb-\zb_n}\,,
\end{eqnarray}
for sources $(q_n^a,\tilde q_n^a)$ at $(z_n,\zb_n)$. 
The first order contribution to the correlation function is then
\be\label{2.11}
\frac{i\lambda\e_{ab}}{(4\pi)^2}\sum_{m,n}[(q_m^a+i\tilde q_m^a)(q_n^b-i\tilde q_n^b)
\int\frac{d^2z}{(z-z_m+\ve)(\zb-\zb_n)}-\mbox{c.c.}]\,.
\ee
The integral is IR divergent. With a cut-off $|z-z_m|<R\gg|z_m-z_n|$, it is
$\sim\pi\log(R/|z_m-z_n|)$ for $m\not=n$ and $\sim\pi\log(R/|\varepsilon|)$ for $m=n$. 
However the $R$-dependence cancels on summing over $m,n$ if we assume overall neutrality of charge and vorticity, so (\ref{2.11}) becomes
\be
-\frac{i\lambda\e_{ab}}{(4\pi)^2}\sum_{m\not=n}[(q_m^a+i\tilde q_m^a)(q_n^b-i\tilde q_n^b)-\mbox{c.c.}]
\pi\log(|z_m-z_n|/|\varepsilon|)
\ee
\be
=\frac{\lambda}{4\pi}\sum_{m\not=n} \e_{ab}\tilde q_m^aq_n^b\log(|z_m-z_n|/|\varepsilon|)\,.
\ee
Note that the correlation function between $m$ and $n$ vanishes if both vorticities are zero. 

It is also worth noting directly from (\ref{2.11}) that the coefficient of the $\log|\ve|$ divergence is
\be\label{2.14}
(\lambda/4\pi)\sum_n\e_{ab}\tilde q_n^aq_n^b\,\log|\ve|\,.
\ee
The origin of this logarithmic divergence may of course be traced to the singular terms in the OPE (\ref{2.7}) with the source fields
\be
\e_{ab}J^a_zJ^b_\zb\sim \e_{ab}\tilde q_n^aq_n^b/z\zb+\cdots\,.
\ee
Note that these terms are prescribed by the Ward identity and therefore exist independently of perturbation theory.

\subsection{Beyond perturbation theory}

We now assume that the original QFT has been deformed by a finite amount and we consider the additional deformation of the correlators of source fields under an infinitesimal change $\lambda\to \lambda+\delta \lambda$. 

\begin{figure}
\centering
\includegraphics[width=0.4\textwidth]{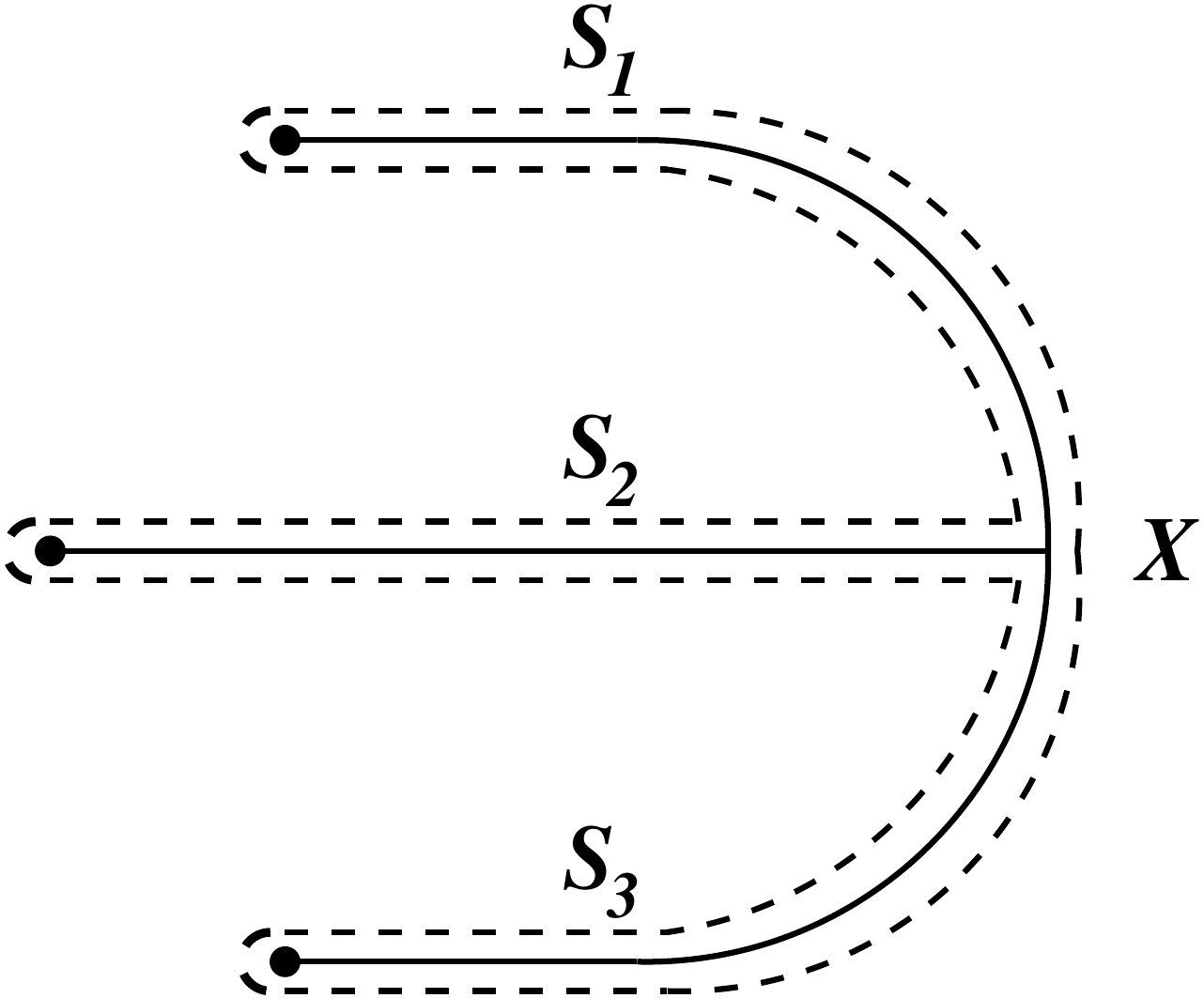}
\caption{\label{Sn} The potential $\chi^b$ is made single-valued by removing the union of paths $S_n$ from each argument $x_n$ to an arbitrarily chosen point $X$.}
\end{figure}

\begin{figure}
\centering
\includegraphics[width=0.4\textwidth]{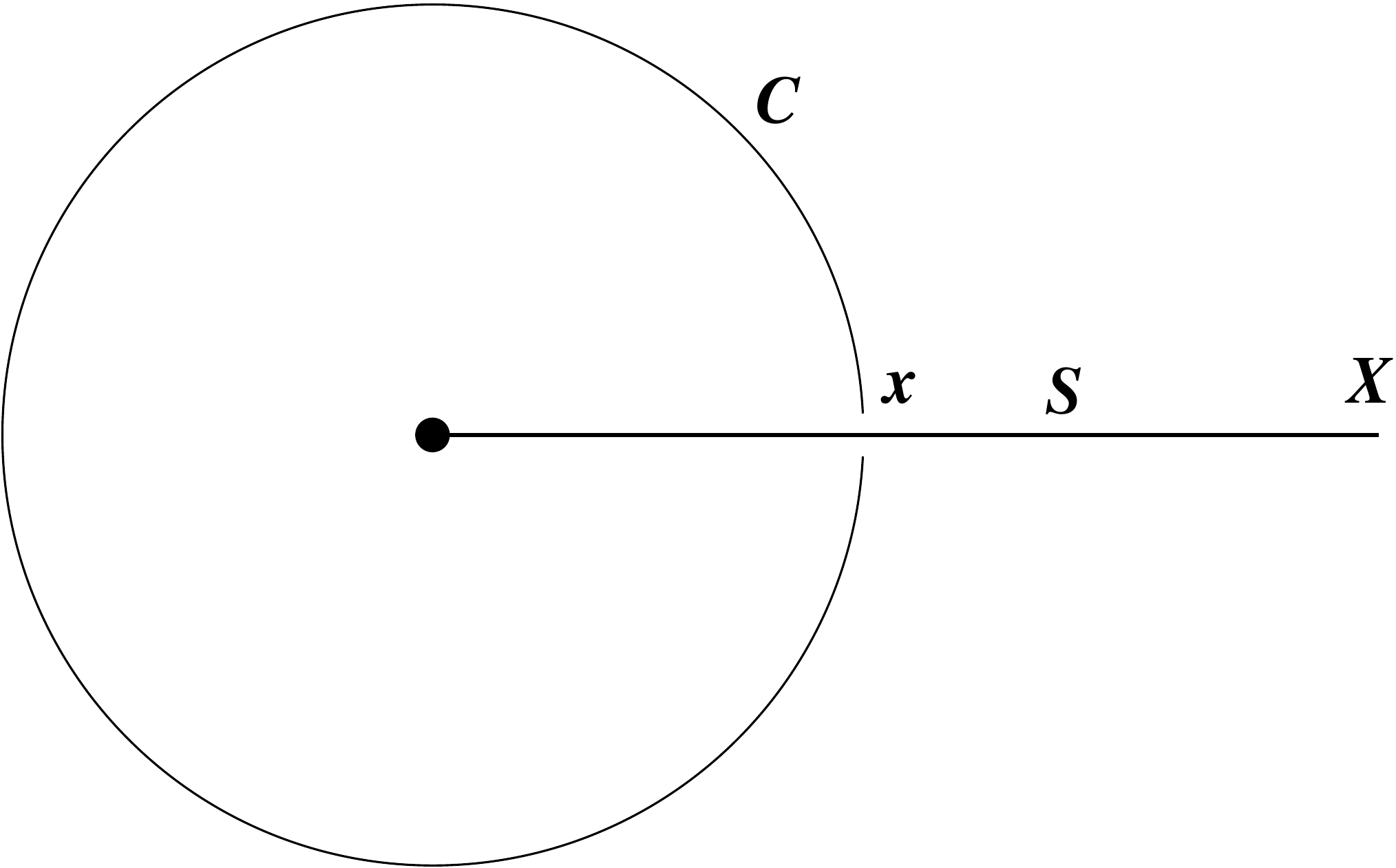}
\caption{\label{contourC} The discontinuity $[\chi^b(x)]$ across $S_n$  is given by the flux of $J^b$ across a contour $C$ surrounding $x_n$.}
\end{figure}

As before, except near the sources or where the fields $\chi^a$ have discontinuities, we may write the deformation in the form 
\be\label{2.16}
\e_{ab}\int J^a_i(x+\ve)\p^i\chi^b(x)d^2x=\e_{ab}\int \p^i[J^a_i(x+\ve)\chi^b(x)]d^2x-\e_{ab}\int \p^i[J^a_i(x+\ve)]\chi^b(x)d^2x\,,
\ee
but we should recall that $\chi^b(x)$ is not single-valued if there is non-zero circulation around the sources.
In order to deal with this 
we insert non-intersecting curves $S_n$ from each source $x_n$ to a prescribed point $X$ (with $|X-x_n|\gg|\ve|$) and restrict the integration to ${\mathbb R}^2\setminus\cup_nS_n$ (see Fig.~\ref{Sn}). 
The first term in (\ref{2.16}) then integrates to
\be
-\sum_n\e_{ab}\int_{S_n}J^a_i(x+\ve)[\chi^b(x)]\e^{ij}dx_j\,,
\ee
where $[\chi^b(x)]$ is the discontinuity in $\chi^b$ across $S_n$. This, in turn, may be written
\be
[\chi^b(x)]=\oint_{C_n(x)}\e^{kl}J^b_k(x)dx_l\,,
\ee
where $C_n(x)$ is a contour beginning and ending at $x$ on $S_n$ and surrounding $x_n$  (Fig.~\ref{contourC}).
Because $J^b$ is conserved, this is independent of the precise contour, and in fact it simply measures the total $b$-charge inside $C_n(x)$.  
Since $J^a$ is assumed to commute with $J^b$, this charge is just that of the source $q_n^b$, giving
\be\label{2.19}
-\e_{ab}q^b_n\int_{S_n}J^a_i(x+\ve)\e^{ij}dx_j\,.
\ee

The second term in (\ref{2.16}) is proportional to $\e_{ab}q^a_n\chi^b(x_n-\ve)$. Again, $\chi^b(x_n-\ve)$ may be written in terms of a contour integral of $J^b$ around $x_n$, which, as $\ve\to0$ is given by the flux $q^b$. Thus this term is
$\propto\e_{ab}q^aq^b=0$. 

The whole contribution of $S_n$ is therefore given by (\ref{2.19}), where the line integral is simply the flux of $J^a$ across $S_n$.
Because $J^a$ is conserved, the integral is independent of the path of $S_n$, and the sum over $n$ is independent of $X$, assuming total charge neutrality. 

However this integral is in general a  non-trivial fluctuating quantity.  From (\ref{2.7}) it diverges as $\ve\to0$  like 
$\sim(1/2\pi)\tilde q^a\log|\ve|$, and we  expect the remainder to be finite, but only if $J^a$ is also curl-free (as happens in a CFT) is its value determined by the near-field circulation. Thus the contribution from $S_n$ is
\be
-(1/2\pi)\e_{ab}q_n^b\tilde q_n^a\log\ve+\mbox{finite as $\ve\to0$}\,.
\ee 
The full result for the infinitesimal deformation is therefore of the form
\be
(\delta\lambda)\sum_n\e_{ab}q_n^b\times\langle\mbox{flux of $J^a$ across $S_n$}\rangle\,.
\ee
If we shift $X\to X'$, the change is proportional to $\sum_nq_n^b\times\mbox{flux of $J^a$ between $X$ and $X'$}$
and therefore vanishes if we have overall charge neutrality. Moreover the paths of the strings may be distorted so as to cross a source, for each time that happens there is an extra contribution $\propto\e_{ab}q_n^b\oint_{\Gamma_n}J^a_i(x')dn'^i=\e_{ab}q_n^aq_n^b$, which vanishes.

For two equal and opposite sources at $x_1,x_2$
\be
(\delta\lambda)\e_{ab}q^b\langle\mbox{flux of $J^a$ across $(x_1+\ve,x_2-\ve)$}\rangle\,.
\ee
The leading logarithmic divergence is determined by the behavior of $\langle J^a(x)\rangle$ as $x\to x_1,x_2$, and is therefore $\propto \tilde q^a$, in agreement with the perturbative calculation (\ref{2.14}), but the $O(1)$ term may depend on less universal details, as well as the deformation parameter $\lambda$. However, 
since this is dimensionless in this case, for a deformed CFT the 2-point function can only depend on the separation $r=|x_1-x_2|$ through the ratio $r/\ve$, so therefore has a universal $\log(r/\ve)$ leading term. Moreover the possibly $\lambda$-dependent remainder is itself proportional to $\e_{ab}q^b$.

\section{$T\overline T$ deformation}\label{sec:TT}

The infinitesimal ``$T\overline T$" deformation is, in Cartesian coordinates, defined  by adding
\footnote{The sign and factors of 2 are chosen so that $\lambda/\pi$ is the same as $-\alpha$ of \cite{Zam1,Zam2} and $-t$ of \cite{JC1}. Thus for $\lambda>0$ the theory exhibits a Hagedorn behavior in its high energy density of states, while $\lambda<0$ corresponds to `going into the bulk' in AdS$_3$.} 
\be
4\pi\delta\lambda\int\det Td^2x=2\pi\delta\lambda\int\e^{ab}\e^{ij}T_{ai}(x+\ve)T_{bj}(x)d^2x
\ee
to the action,
which is formally the same as the $J^1\wedge J^2$ deformation with the identification $J^a_i\to T_{ai}$.
However there is an important difference in that $T_{ai}$ is a rank 2 current with dimension 2, so the deformation parameter $\lambda$ has dimension (length)$^2$, and the corresponding `charges' transform as vectors. 

\subsection{First-order deformation about a CFT}\label{sec:CFT}
Again it is instructive to consider first the perturbative result to first order in a CFT. In complex coordinates the deformation of the action is
$-16\pi\lambda \int T(z+\ve)\overline T(\zb)d^2z$. The first order correction to a CFT correlator $\langle\prod_p\Phi_p(x_p)\rangle$
is therefore
\be
16\pi\lambda\int\langle T(z+\ve)\overline T(\zb) \prod_p\Phi_p(z_p,\zb_p)\rangle d^2z\,.
\ee
We assume that $|z_m-z_n|>\ve$ if $m\not=n$, but make no assumption on the conformal spins of the fields, or whether they are primaries.

This is given by the conformal Ward identity as\footnote{We do not incorporate the conventional factor $2\pi$ in the definition of $T_{ji}$.}
\be
\frac{16\pi\lambda}{(2\pi)^2}\sum_{m,n}\int\sum_{r,s\geq1}\frac{d^2z}{(z-z_m+\ve)^r(\zb-\zb_n)^s}\left\langle (L_{r-2}\Phi_m)(\overline L_{s-2}\Phi_n)\prod_{p\not=m,n}\Phi_p(z_p,\zb_p)\right\rangle\,.
\ee
The integral in the term $r=s=1$, proportional to $\p_{z_m}\p_{\zb_n}\langle\prod_p\Phi_p\rangle$, was already encountered in Sec.~\ref{sec:JJ}, and is 
$\pi\log(R/|\ve|)$ if $m=n$ and  $\pi\log(R/|z_m-z_n|)$ otherwise. These terms then sum to
\be
-(4\lambda)\sum_{m\not=n}\log(|z_m-z_n|/|\ve|)\p_{z_m}\p_{\zb_n}\langle\prod_p\Phi_p(z_p,\zb_p)\rangle\,,
\ee
where we have used translational invariance $\sum_m\p_{z_m}\langle\prod_p\Phi_p\rangle
=\sum_n\p_{\zb_n}\langle\prod_p\Phi_p\rangle=0$. 

In fact 
all the other terms are zero. Those with $r\geq2$ and $s\geq2$ are proportional to derivatives $\p_{z_m}^{r-1}\p_{\zb_n}^{s-1}$ of the $r=s=1$ integral and therefore vanish. If $r=1$, $s\geq2$
 we may shift the integration variable
giving
\be
\int\frac{d^2z}{z(\zb-\zb_{nm})^s}\,,
\ee
where $\zb_{nm}=\zb_n-\zb_m-\bar\ve$. For $|z|<|z_{nm}|$ the integrand may be expanded in powers $\zb^k/z\zb_{nm}^{k+s}$, and for $|z|>|z_{nm}|$ in powers of $\zb_{nm}^k/z\zb^{k+s}$, with $k\geq0$. But all these terms vanish on angular integration.\footnote{The above manipulations are delicate since the integrals are not absolutely convergent. This may however be addressed by introducing a second UV cutoff $|z-z_m|>\ve'$, $|\zb-\zb_n|>\ve'$ with $\ve'\ll\ve$.} Similarly if $r\geq2$, $s=1$.

We conclude that the  first order correction to the CFT correlation function is
\be\label{3.7}
-(4\lambda)\left( \sum_{m\not=n}\log(|z_m-z_n|/|\ve|)\p_{z_m}\p_{\zb_n}\right)\langle\prod_p\Phi_p(z_p,\zb_p)\rangle+O(\ve)\\,
\ee
or, in Cartesian coordinates,
\be
-\lambda\left( \sum_{m\not=n}\log(|x_m-x_n|/|\ve|)\p^{x^i_m}\p_{x^i_n}\right)\langle\prod_p\Phi_p(x_p)\rangle\,.
\ee

If we want to isolate only the $\ve$-dependence, this is
\be\label{3.9}
\lambda(\log|\ve|)\sum_{m\not=n}\p_{x^i_m}\p_{x^i_n}\langle\prod_p\Phi_p(x_p)\rangle
=-\lambda(\log|\ve|)\sum_{n}\nabla^2_{x_n}\langle\prod_p\Phi_p(x_p)\rangle\,,
\ee
using translational invariance.
Note that this comes from the $O(1/z\zb)$ term in the OPE and so is universal, as we shall see in the next section. 
Also, if any of the $\Phi_p$ is holomorphic (or antiholomorphic), as for a conserved current in a CFT,  then its first-order variation vanishes identically.

Finally we record the result for the two-point function 
$C^\lambda(x)=\langle \Phi(x_1)\Phi(x_1+x)\rangle$, as also found in \cite{Kraus}:
\be\label{2ptx}
C^\lambda(x)=\left(1+2\lambda\log(|x/\ve|)\,\nabla^2_x+O(\lambda^2)\right)C^0(x)\,.
\ee

\subsection{Beyond perturbation theory}
The analysis of the $J^1\wedge J^2$ deformation may be translated almost line by line, with the identification
$J^a_i\to T^a_i$, with an important difference: since 
\be
\p_{x_i}T^b_i(x)\Phi_n(x_n)=\d^{(2)}(x-x')\p_{x_n^b}\Phi_n(x_n)\,,
\ee
the ``charge'' $q^b$ associated with $T^b$ is a spatial derivative $\p^b$, and all fields, including $T^a$, carry charge under this symmetry current.

We may however still take over the results of Sec.~\ref{sec:JJ} to argue that the bare evolution equation for a generic correlation function has the form
\be\label{3.11d}
\p_\lambda\langle\prod_p\Phi_p(x_p)\rangle_{\lambda}=
2\pi\sum_n\e^{ab}\e^{ij}\int_{x_n}^Xdx'_j\langle T^{\lambda}_{ai}(x'+\ve)\,\p_{x_n^b}\prod_p\Phi_p(x_p)\rangle_{\lambda}\,.
\ee
The solution of this equation is
\be\label{3.12e}
\langle\prod_p\Phi_p(x_p)\rangle_{\lambda}=\langle\prod_p\Phi^\lambda_p(x_p)\rangle_{\lambda}\,,
\ee
where
\be\label{3.14f}
\p_\lambda\langle\prod_p\Phi^\lambda_p(x_p)\rangle_{\lambda}=\sum_n\langle\p_\lambda\Phi^\lambda_n(x_n)
\prod_{p\not=n}\Phi^\lambda_p(x_p)\rangle_{\lambda}\,,
\ee
and
\be\label{3.13}
\p_\lambda\Phi^\lambda_n(x_n)
=2\pi\e^{ab}\e^{ij}\int_{x_n}^Xdx'_j {T^\lambda}_{ai}(x'+\ve)\,\p_{x_n^b}\Phi^\lambda_n(x_n)\,,
\ee
with $\Phi^{\lambda=0}_n(x_n)=\Phi_n(x_n)$.

Although by conservation (\ref{3.13}) is invariant under local deformations of the path $S_n$ connecting $x_n$ to $X$, it is also unchanged if, say, $S_n$ is deformed through some other argument $x_m$. For then the residual extra term is of the form $\e^{ab}\oint\oint T_{ai}(x')T_{bj}(x'')dn'^idn''^j$, where the contours surround $x_m$, and this vanishes by antisymmetry. Similarly (\ref{3.11d}) is unchanged if some of the paths happen to cross one another.
The only constraint is that the arguments should not actually lie on a path. 
This is another consequence of the topological nature of the $\det T$ deformation. 

In addition, if we act with the derivative $\p_{X_l}$ on the sum over $n$ in (\ref{3.11d}) we get something proportional to
\be
\e^{ab}\e^{il}{T^\lambda}_{ai}(X+\ve)\sum_n\p_{x_n^b}\prod_{p}\Phi^\lambda_p(x_p)
=\e^{ab}\e^{il}{T^\lambda}_{ai}(X+\ve)\sum_n\oint_{x_n}T^\lambda_{bk}dn^k\prod_{p}\Phi^\lambda_p(x_p)\,.
\ee
The contours around each argument may be distorted to one around a large circle at infinity, which vanishes in a translationally invariant state, plus one around $X$, which vanishes by the same antisymmetry argument as above.

Eq.~(\ref{3.13}) is the main result of this paper. It shows that fields which are local in the undeformed theory evolve into semilocal fields, very similar to disorder or twist fields in conventional QFT. 
That the evolution of the correlation function may be expressed simply in terms of an evolution of each field is a non-trivial consequence of the partial solvability of the $\det T$ deformation, which allows the insertion of $\int\det T d^2x$ to be written as a sum over an integral along $S_n$ times one around each argument $x_n$. 
In particular (\ref{3.14f}) shows that the deformation acts on a product of fields by the Leibniz rule, and is therefore a derivation on the algebra of such fields. 
We stress, however, that  $T^{\lambda}$ is to be evaluated in the $\lambda$-deformed theory, and that it describes the bare evolution of fields $\Phi_n$, in the sense that it is assumed that  their definition is $\lambda$-independent. 
In particular, if we define the correlators as functional derivatives with respect to sources, these source terms should be added into the deformed action, not to the undeformed theory, because they would then enter into the definition of $\det T^\lambda$.
In the case of conserved currents, including the stress tensor itself, there are additional terms coming from the explicit change in the action, to be discussed below.

\subsubsection{Example: 2-point function}
Since the general result (\ref{3.13}) is rather telegraphic, it is worth exhibiting how it applies to the simplest case of a two-point function. Taking the two arguments to lie on $x^0=0$, (\ref{3.14f}) and (\ref{3.13}) give, in Cartesian coordinates 
\be
\p_\lambda\langle\Phi(0,x_1)\Phi(0,y_1)\rangle_\lambda
=2\pi\e_{ab}\e_{i1}\left(\int_{x_1+\ve}^Xdx'_1\p_{x_b}+\int_{y_1+\ve}^Xdx'_1\p_{y_b}\right)\langle T_{ai}(0,x'_1)
\Phi(0,x_1)\Phi(0,y_1)\rangle_\lambda
\ee
\be
=2\pi\e_{ab}\int_{x_1+\ve}^{y_1-\ve}dx'_1(\p_{x_b}-\p_{y_b})\langle T_{a0}(0,x'_1)
\Phi(0,x_1)\Phi(0,y_1)\rangle_\lambda\,,
\ee
using translational invariance. The term with $(ab)=(10)$ vanishes by symmetry, and we are left with
\begin{eqnarray}
\!\!\!\!\!\!\!\!\!\!\!\!\!\!\!\!\p_\lambda\langle\Phi(0,x_1)\Phi(0,y_1)\rangle_\lambda
&=&2\pi\int_{x_1+\ve}^{y_1-\ve}dx'_1(\p_{x_1}-\p_{y_1})\langle T_{00}(0,x'_1)
\Phi(0,x_1)\Phi(0,y_1)\rangle_\lambda\label{4.14e}\,,\\
&-&2\pi\int_{x_1+\ve}^{y_1-\ve}dx'_1(\p_{x_0}-\p_{y_0})|_{x_0,y_0=0}\langle T_{10}(0,x'_1)
\Phi(x_0,x_1)\Phi(y_0,y_1)\rangle_\lambda\label{4.15e}\,.
\end{eqnarray}
The insertions $\int_{x_1+\ve}^{y_1-\ve}T_{a0}(0,x'_1)dx'_1 $ are the energy and momentum circulation between the two points (cut off close to them), and (\ref{4.14e}, \ref{4.15e}) measure how this is correlated with moving them an infinitesimal distance apart. Alternatively,  they give the response of the separation of the two points to the insertion of an infinitesimal time delay and boost between them. This interpretation will become more clear in Sec.~\ref{sec:diffeo}.

\subsection{Deformation of conserved currents}

We now see how the above argument should be  modified when the field $\Phi(x)$ is a conserved current $J_c(x)$. 
In considering deformed symmetry currents, there are three important issues to address:  does the deformed current remain conserved (which it should if the deformation does not break the symmetry); is it finite (apart from possible total derivatives); and does it still have the correct OPE with charged fields, as dictated by its Ward identity?
We proceed inductively, that is, assume that $J_c^\lambda$ has the desired properties, and ask whether these continue to hold for $J_c^\lambda+(\d\lambda)\p_\lambda J_c^\lambda$.

First consider a conserved vector current $J^\lambda_c$.  We have from (\ref{3.13}) that its deformation due to the insertion of $\int\det T^\lambda \,d^2x$ is given by
\be\label{3.32e}
\p_\lambda J_c^\lambda(x)
=2\pi\e^{ab}\e^{ij}\int_x^Xdx'_jT^\lambda_{ai}(x'+\ve)\,\p_{x^b}J_c^\lambda(x)\,.
\ee
As it stands, even if $J^\lambda_c$ is conserved, its deformation is not, $\p_\lambda \p_{x_c}J_c^\lambda(x)\not=0$, since $\p_{x_c}$ also acts on the lower limit of integration. In fact
\be
 \p_\lambda \p^cJ_c^\lambda(x)=-2\pi\e^{ab}\e^{ic}T^\lambda_{ai}(x+\ve)\p_{b}J_c^\lambda(x)
 \ee
 \be
=2\pi{T^\lambda}^a_i(x+\ve)\p^iJ_a^\lambda(x)-2\pi{T^\lambda}^a_a(x+\ve)\p^iJ_i^\lambda(x)=2\pi\p^c[{T^\lambda}^a_c(x+\ve)J_a^\lambda(x)]\,.
 \ee
 This result may be checked by directly computing the OPE of the point split version of $\det T$ with $J_c$.
 
  Thus we should incorporate this into the right hand side of (\ref{3.32e}) to define the conserved infinitesimally deformed current by
\be\label{3.39e}
D_\lambda J_c^\lambda(x)=2\pi\e^{ab}\e^{ij}\int_x^Xdx'_jT^\lambda_{ai}(x'+\ve)\p_bJ_c^\lambda(x)
-2\pi{T^\lambda}^a_c(x+\ve)J_a^\lambda(x)\,,
\ee
which, using the identity $\e^{ab}J_c=\e^{cb}J_a+\e^{ac}J_b$, is
\be\label{3.41e}
2\pi\e^{cb}\e^{ij}\int_x^Xdx'_jT^\lambda_{ai}(x'+\ve)\p_bJ_a^\lambda(x)
-2\pi{T^\lambda}^a_c(x+\ve)J_a^\lambda(x)
=2\pi\e^{cb}\e^{ij}\p_{x^b}\int_x^Xdx'_jT^\lambda_{ai}(x'+\ve)J_a^\lambda(x)\,.
\ee
This last form shows explicitly that $D_\lambda \p^cJ_c^\lambda(x)=0$ even in the presence of sources:  the deformation of the current is purely transverse, affecting only its circulation and not its divergence. Thus it continues to satisfy the appropriate Ward identity, and the total charge, given by space integral of $J^\lambda_0$, is not deformed.

Now consider the dependence of $D_\lambda J_c^\lambda(x)$ on the cut-off $\ve$. A simple argument shows that $D_\lambda J_c^\lambda(x)$ is in fact UV finite.
Acting with $\p_{\ve^l}$ on the right hand side of (\ref{3.41e}) gives something proportional to
\be\label{3.24d}
\e^{cb}\e^{il}\p_{x^b}[T^\lambda_{ai}(x+\ve)J_a^\lambda(x)]\,.
\ee
In principle this OPE is singular. However, assuming that $T^\lambda_{ai}$ is neutral under $J^\lambda_c$ (for example, it is an internal rather than a space-time symmetry), the OPE between them, expanding in a basis of fields depending on $x+\ve$ rather than on $x$, should be non-singular:
\be
J^\lambda_a(x)\cdot T^\lambda_{ai}(x+\ve)=:\!\!J_aT_{ai}\!\!:^\lambda\!\!(x+\ve)+o(|\ve|^0)\,,
\ee
so that, if $J_c^\lambda(x)$ is finite as $\ve\to0$, so is its infinitesimal deformation $\p_\lambda J_c^\lambda(x)$.
Thus the deformation of a conserved symmetry current, when properly defined, is both finite and satisfies the Ward identities.

\subsubsection{Deformation of the stress tensor}\label{sec:Tdef}

Turning to the case of the stress tensor $T^\lambda_{dc}$ itself, much of the above argument may be carried over.
The analogs of (\ref{3.39e}, \ref{3.41e}) are
\begin{eqnarray}
D_\lambda T_{dc}^\lambda(x)&=&2\pi\e^{ab}\e^{ij}\int_x^Xdx'_jT^\lambda_{ai}(x'+\ve)\p_bT_{dc}^\lambda(x)
-2\pi{T^\lambda}^a_c(x+\ve)T_{da}^\lambda(x)\label{3.39f}\,,\\
&=&2\pi\e^{cb}\e^{ij}\p_{x^b}\int_x^Xdx'_j{T^\lambda}^a_i(x'+\ve)T_{da}^\lambda(x)\label{3.40e}\,.
\end{eqnarray}
Note that if we use the more symmetric form (\ref{symm}) of point splitting, then from (\ref{3.39f}) if $T_{dc}^\lambda$ is symmetric in its indices, as expected for a Lorentz invariant theory, then so is 
$D_\lambda T_{dc}^\lambda$, although this is hidden in (\ref{3.40e}). Also, in complex coordinates, the last term in
(\ref{3.39f}) adds a term $\propto T\overline T$ to the deformation of the trace, as expected since the action itself is deformed \cite{Kraus}. Once again, the form in (\ref{3.40e}) ensures that $T_{dc}^\lambda$ continues to satisfy its Ward identity. 

However, the argument above that the deformation of a conserved current that commutes with $T_{ai}^\lambda$
is UV finite does not carry through, since $T_{ai}^\lambda$ is itself charged under $T_{dc}^\lambda$. 
Indeed, if we take the 
the derivative with respect to $\ve_l$ of (\ref{3.40e}) we find something proportional to
\be
\p^{x^b}\e_{cb}\e^{il}[{T^\lambda}^a_i(x+\ve)T_{da}^\lambda(x)]\,,
\ee
where, since because we should symmetrize and therefore need only the piece which is odd, this is given by the
Ward identity term in the OPE
\be
{T^\lambda}^a_i(x+\ve)T_{da}^\lambda(x)=(1/\ve^2)\left(\ve_i\p_a-\e^{af}\e^{ik}\ve_k\p_f\right)T_{da}\,,
\ee
giving
\be\label{Tdiv}
(1/\ve^2)\p^{x^b}\e_{cb}\e^{il}\e^{af}\e^{ik}\ve_k\p_fT_{da}=(\ve_l/\ve^2)\p^b\p_bT_{dc}\,.
\ee
On integration, this gives a $\log|\ve|$ divergence which, as we shall see in the next section, afflicts all fields of the deformed theory. However, consistently with (\ref{3.40e}), this does not affect its Ward identity, and can if wished  be subtracted off as a further improvement, since it is a total derivative. This point of view will be expanded in Sec.~\ref{sec:renorm}.

\section{Analysis of divergences}\label{sec:div}
Let us rewrite Eq.~(\ref{3.13}) as
\be\label{4.1d}
\p_\lambda\Phi^\lambda(x)
=2\pi\e^{ab}\e^{ij}\int_{x+\ve}^Xdx'_j {T^\lambda}_{ai}(x')\,\p_{x^b}\Phi^\lambda(x)\,,
\ee
This in general exhibits explicit UV divergences as $\ve\to0$. These may be analyzed by using the OPE with $T^\lambda$. Although we have shown that $T^\lambda$ may be deformed in such a way that it has the correct properties, (\ref{4.1d}) also contains implicit divergences in $\Phi^\lambda$ on the right hand side. However its nested form allows us to treat these iteratively. The OPE of $T^\lambda_{ai}(x')$ with $\p_{x^b}\Phi(x)$ in principle contains arbitrarily high order terms 
$O(\lambda^N/|x'-x|^{k+2N})$, and those with $k+2N>1$ lead to power law divergences which may proliferate as with any non-renormalizable deformation. However we now argue that such terms do not contribute to (\ref{4.1d}), just as for the first order CFT calculation in Sec.~(\ref{sec:CFT}). 

Taking $\p_{\ve_k}$ to expose any divergences
\be\label{4.2d}
-2\pi\e^{ab}\e^{ik} {T^\lambda}_{ai}(x+\ve)\,\p_{x^b}\Phi^\lambda(x)\,.
\ee
However (\ref{4.1d}) is independent of the direction of $\ve_k$, so (\ref{4.2d}) is parallel to $\ve_k$ and we may average uniformly over it to obtain
\be
-|\ve|^{-1}\oint d\ve_k\e^{ab}\e^{ik} {T^\lambda}_{ai}(x+\ve)\,\p_{x^b}\Phi^\lambda(x)
\ee
\be
=|\ve|^{-1}\oint d\ve^a {T^\lambda}_a^b(x+\ve)\,\p_{x^b}\Phi^\lambda(x)
-|\ve|^{-1}\oint d\ve^b {T^\lambda}_a^a(x+\ve)\,\p_{x^b}\Phi^\lambda(x)
\ee
The first term gives, by the Ward identity, $|\ve|^{-1}\p_{x_b}\p_{x^b}\Phi^\lambda(x)$, in agreement with the first-order CFT calculation in Sec.~(\ref{sec:CFT}) on integrating back with respect to $|\ve|$. As far as any singular terms are concerned, the second term is equivalent to
\be
-|\ve|^{-1}\oint d\ve^b {T^\lambda}_{aa}(x)\,\p_{x^b}\Phi^\lambda(x-\ve)
=|\ve|^{-1}\oint d\ve^b {T^\lambda}_{aa}(x)\,\p_{\ve^b}\Phi^\lambda(x-\ve)\,,
\ee
which vanishes.

A more explicit way of deriving the logarithmic divergence in (\ref{4.1d}) is as follows.
 The relevant terms in the OPE are completely determined to be of the form 
\be\label{3.13d}
(2\pi)T^\lambda_{bj}(x)\,\Phi(0)=(x_j/|x|^2)\p_b\Phi(0)+\xi(\e_{jk}x^k/|x|^2)\e_{ba}\p^a\Phi(0)+\cdots\,,
\ee
where the first term is fixed by the Ward identity $\p^jT^\lambda_{bj}(x)\Phi(0)=\d^{(2)}(x)\p_{b}\Phi(0)$, and the form of the remainder, which is orthogonal to $x_j$, is fixed by rotational symmetry and parity. Conservation, or symmetry under $b\leftrightarrow j$, then fixes $\xi=-1$. In the language of Sec.~\ref{sec:JJ}, the first term gives the ``charge'', which is now a vector field $q\sim \p_b$, and the second term gives the vorticity $\tilde q\sim-\e_{ba}\p^a$. Note that the contribution from the trace $T^j_j$ vanishes on the right hand side, and in fact these terms in the OPE have the same form as in a CFT, a peculiar property of two-dimensional field theories.

Inserting the OPE (\ref{3.13d}) into the integral along $S_n$ we find
 \be\label{3.18}
 \p_\lambda\Phi^\lambda(x)=-(\log|\ve|)\nabla^2_{x}\Phi^\lambda(x)+\cdots\,.
  \ee
 Again, the leading term in (\ref{3.18}) is in agreement  
with the perturbative calculation (\ref{3.9}). The remainder 
  is in general different, but is finite when expressed in terms of $\Phi^\lambda$.
   
At this point, there are two paths we may take: either regard the $\Phi^{\lambda=0}_n$ as the physical fields of the effective theory, and try to resum the UV divergences in their correlation functions to obtain closed form results; or to redefine the evolution of the $\Phi^\lambda_n$ so that their correlators are finite to all orders. 

\subsection{Resummation of leading logs}

 The iterative structure of (\ref{3.13}) allows us to make the following argument. 
 Together with (\ref{3.18}),  this shows that if we expand an arbitrary correlation function  in powers of $\lambda$, the structure is
\be
\langle\prod_n\Phi_n(x_n)\rangle_\lambda=\sum_{N=0}^\infty (\lambda^N/N!)(A_N(\{x_n\})(-\log|\ve|)^N+B_N(\{x_n\})(-\log|\ve|)^{N-1}+\cdots)\,,
\ee
where
\be
A_N(\{x_n\})=\sum_n\nabla^2_{x_n}A_{N-1}(\{x_n\})\,,
\ee
\be
B_N(\{x_n\})=\sum_n\nabla^2_{x_n}B_{N-1}(\{x_n\})+\cdots\,,
\ee
so that
\be
\langle\prod_n\Phi_n(x_n)\rangle_\lambda=\sum_{N=0}^\infty (-\lambda\log|\ve|)^N/N!)(A_N(\{x_n\})-\lambda B_{N+1}(\{x_n\})+\cdots)\,.
\ee
Apart from the factor $\nabla^2_{x_n}$ this very like what one would see in a locally renormalizable theory, except that
$\lambda$ retains its canonical dimension. Therefore we may introduce a length scale $\mu^{-1}$ and take a  scaling limit
\be
\ve\mu\to0\,,\quad \lambda\mu^2\to0\quad\mbox{with $\tilde\lambda\mu^2=-\lambda\mu^2\log|\ve\mu|$ fixed,}
\ee
in which only the leading terms $A_N$ survive, and the correlation functions exactly satisfy 
 a $2n$-dimensional diffusion equation with respect to their arguments. The solution is then
\be\label{3.51d}
\langle\prod_n\Phi_n(x_n)\rangle_\lambda=\int \prod_n G(x_n-y_n;\tilde\lambda)\langle\prod_n\Phi_n(y_n)\rangle_0\prod_nd^2y_n\,,
\ee
where
\be
G(x-y;\tilde\lambda) =(4\pi\tilde\lambda)^{-1}e^{-(x-y)^2/4\tilde\lambda}
\ee
is the 2d heat kernel. Note that further subtractions would need to be made, using the OPE of the undeformed theory, if there are non-integrable singularities in the above equation. 

However, at least in an infinite system, this diffusive behavior makes sense only for $\tilde\lambda>0$, that is $\lambda>0$ (the sign corresponding to Hagedorn behavior.) For the `wrong' sign, the IR behavior immediately diverges.

\subsection{Beyond leading logs: renormalization}\label{sec:renorm}
An alternative point of view is to try to define deformed fields whose correlators are finite to all orders, similar in spirit to the renormalization program in a conventional local renormalizable field theory. Returning to (\ref{3.13}) 
\be
\p_\lambda\Phi^\lambda(x)
=2\pi\e^{ab}\e^{ij}\int_{x}^Xdx'_j {T^\lambda}_{ai}(x'+\ve)\,\p_{x^b}\Phi^\lambda(x)\,,
\ee
we may regard this as a Schr\"odinger-like equation (without the $i$) with $\lambda$ playing the role of time, and $2\pi\e^{ab}\e^{ij}\int_{x}^Xdx'_j {T^\lambda}_{ai}(x'+\ve)\,\p_{x^b}$ being a time-dependent `hamiltonian' acting on the vector space of fields of the theory (in a CFT this would be a Virasoro module).

In the leading log approximation, 
\be
\p_\lambda\Phi^\lambda(x)
\approx -\log|\ve|\nabla^2_x\Phi^\lambda(x)\,,
\ee
so that $\Phi^\lambda(x)
\approx e^{-\lambda\log|\ve|\nabla^2_x}\Phi^0(x)$.
This suggests that we take into account the corrections by going to an `interaction picture', defining
\be
 \widehat\Phi^\lambda(x)
\equiv e^{\lambda\log|\mu\ve|\nabla^2_x}\,\Phi^\lambda(x)\,,
\ee
and similarly\footnote{In this picture the stress tensor plays a dual role, as the kernel of the evolution operator, and as a field acted on by this operator. Of course, this is familiar from conventional hamiltonian dynamics.}
\be
 \widehat T^\lambda_{ai}(x'+\ve)
\equiv e^{\lambda\log|\mu\ve|\nabla^2_{x'}}\,
T^\lambda_{ai}(x'+\ve)\,e^{-\lambda\log|\mu\ve|\nabla^2_{x'}}\,.
\ee
Here $\mu$ is an arbitrary renormalization scale with dimensions of inverse length.
Note that $e^{\pm\lambda\log|\mu\ve|\nabla^2_x}$ commutes with $\int_xdx'_j$ and $\p_{x^b}$, and that, by (\ref{Tdiv}), this kills the leading log divergences in $T^\lambda_{ai}$ in the same manner as for those in $\Phi^\lambda$.  

$\widehat\Phi^\lambda(x)$ then satisfies
\be\label{3.57}
\p_\lambda \widehat\Phi^\lambda(x)
=2\pi\e^{ab}\e^{ij}\int_{x}^Xdx'_j  \widehat T^\lambda_{ai}(x'+\ve)\,\p_{x^b}      \widehat\Phi^\lambda(x)+\log|\mu\ve|\nabla^2_x\widehat\Phi^\lambda(x)\,.
\ee
Now the $O(1/|x'-x))$  terms in the OPE of $\widehat T^\lambda_{ai}(x')$ with
$\Phi^\lambda(x)$ are the \em same \em as those of $T^\lambda_{ai}(x')$, and are given by the Ward identity (\ref{Tdiv}). This is because these terms are all of the form
$\p_{x'}\log|x'-x|$, and are annihilated by $\nabla^2_{x'}$. Thus the last term in 
(\ref{3.57}) exactly cancels this divergence, and if the correlators of $\widehat\Phi^\lambda(x)$ are finite, so are those of 
$\p_\lambda \widehat\Phi^\lambda(x)$. 

We have therefore shown, generalizing (\ref{3.51d}), that, to all orders 
\be
\langle\prod_n\Phi_n(x_n)\rangle_\lambda=\int \prod_n G(x_n-y_n;\lambda)\langle\prod_n\widehat\Phi_n(y_n)\rangle_\lambda\prod_nd^2y_n\,,
\ee
where
\be
G(x-y;\lambda) =(4\pi\lambda|\log|\mu\ve||)^{-1}e^{-(x-y)^2/\lambda|\log|\mu\ve||}
\ee
and $\langle\prod_n\widehat\Phi_n(y_n)\rangle_\lambda$ is finite as we remove the cut-off $\ve\to0$.

As an example, to first order about a deformed CFT, we find from (\ref{3.7}) that
\be\label{3.42c}
\p_\lambda\langle\prod_n\widehat\Phi_n(y_n)\rangle_\lambda=
-\sum_{m\not=n}\log(\mu|x_m-x_n|)\p^{x_m^i}\p_{x_n^i}\langle\prod_n\widehat\Phi_n(y_n)\rangle_{\rm CFT}+O(\lambda)\,.
\ee

\subsection{Renormalization group}\label{sec:RG}
The fact that the bare correlation functions $\langle\prod_n\Phi_j(x_j)\rangle_\lambda$ do not depend on the renormalization scale $\mu$, allows us, as usual, to infer a Callan-Symanzik equation, most simply written in terms of the Fourier transform in $k$-space
\be
0=\mu\p_\mu C(\{k_n\};\lambda,\ve)=\mu\p_\mu\left[\prod_ne^{\lambda\log(\mu\ve) k_n^2}\,\widehat
C(\{k_n\};\lambda,\mu)\right]\,,
\ee
so
\be
[\mu\p_\mu+\lambda\sum_nk^2_n]\widehat
C(\{k_n\};\lambda,\mu)=0\,.
\ee

Specializing to the two-point functions of a deformed massless theory, where
$C(k;\lambda=0)\sim k^{2\Delta}$, by dimensional analysis, 
\be
[\mu\p_\mu+k\p_k-2\lambda\p_\lambda-2\Delta]\widehat
C(k;\lambda,\mu)=0\,,
\ee
so that, at fixed $\mu$
\be
\left[k\p_k-2\lambda\p_\lambda-2\lambda k^2-2\Delta\right]\widehat
C(k;\lambda,\mu)=0\,.
\ee
This linear first order PDE has the solution matching onto $\lambda=0$
\be\label{ck}
\widehat
C(k;\lambda,\mu)=k^{2\Delta}(k/\mu)^{2\lambda k^2}
\ee
 In fact, having made this analysis, we may choose $\mu=|\lambda|^{-1/2}$ as long as we now treat $\lambda$ as fixed away from zero. This gives the form quoted in the abstract.
 
 In real space we then have, at least formally,
 \be
 \widehat
C(x;\lambda,\mu)=\int k^{2\Delta}e^{\lambda k^2\log(k^2/\mu^2)-ik\cdot x}d^2k
\ee
The formal perturbative expansion agrees with (\ref{3.7}) to first order in $\lambda$, and is an  asymptotic expansion valid for $|x|\gg\sqrt\lambda$. The fact that the integral appears to diverge for large real $k$ if $\lambda>0$ may be controlled by suitably distorting the contour in $k$ as $\lambda$ is continued from negative values, just as for the simpler gaussian integral without the log factor. For either sign the interesting  limit is when $|x|\ll\sqrt\lambda$. This limit, and the large order 
behavior of the perturbative expansion, are given by a saddle point of the integral, which occurs at
\be
2\lambda k_c(1+\log(k_c^2/\mu^2))=ix
\ee
There are two solutions, one with $k_c\sim ix/\big(2\lambda\log(-x^2/4\lambda^2\mu^2)\big)$ which gives rise to a behavior
\be
\sim e^{-x^2/\big(4\lambda|\log(x^2/4\lambda^2\mu^2)|\big)}
\ee
times prefactors. The second solution has $\log(k_c^2/\mu^2)\approx -1$ and gives rise to damped oscillatory behavior in $x$ on the scale $\mu^{-1}$, independent of $\lambda$. A more careful analysis reveals that the first solution is appropriate for $\lambda>0$ and the second for $\lambda<0$. It is tempting to associate the first with the Hagedorn growth in the density of states in finite volume, and the second with the phenomenon that all energies become imaginary with the same real part, but to make this identification systematic would require computing the correlation functions on the cylinder.

\subsection{Deformed OPE}
We have argued that the $T\overline T$  deformation acts as a derivation on the algebra of local fields, that is it satisfies the Leibniz rule when applied to correlators of products of fields at distinct points. Although it therefore preserves the fusion algebra of the OPEs, it will in general modify the OPE coefficients and the conformal blocks. Given the short distance OPE in the undeformed theory
 \be
\Phi_m(x_1)\cdot\Phi_n(x_2)=\sum_l{C}^l_{mn}(x_{12})\Phi_l(\bar x)\,,
\ee
(where $\bar x=(x_1+x_2)/2$, $x_{12}=x_1-x_2$) we may ask whether the deformed OPE
\be
\widehat\Phi^\lambda_m(x_1)\cdot\widehat\Phi^\lambda_n(x_2)=\sum_l{C^\lambda}^l_{mn}(x_{12})\widehat\Phi^\lambda_l(\bar x)
\ee
makes sense inside deformed correlators defined by (\ref{3.57}). 

Much of this structure may already be seen  at first order.
Taking $|x_{12}|\ll|x_{1n}|$ for $n\geq3$ in (\ref{3.42c}),
\begin{eqnarray}
&&\p_\lambda|_{\lambda=0}\,\langle\prod_p\widehat\Phi_p(x_p)\rangle_{\lambda}\sim\nonumber\\
&&-\left(\log(|x_{12}|\mu)\p_{x^b_1}\p^{x^b_2}+\sum_{n\geq3}\log(|x_{1n}|\mu)\p_{x^b_1}\p^{x^b_n}
+\sum_{n\geq3}\log(|x_{2n}|\mu)\p_{x^b_2}\p^{x^b_n}+\cdots\right)\nonumber\\
&&\left(\sum_l{C}^l_{12}(x_{12})\langle\Phi_l(\bar x)\prod_{n\geq3}\Phi_n(x_n)\rangle_{\rm CFT}\right)\label{4.3d}\,,
\end{eqnarray}
which should be equated to
\be
\sum_l\left(\p_\lambda[{C}^l_{12}(x_{12})]\langle\Phi_l(\bar x)\prod_{n\geq3}\Phi_n(x_n)\rangle
+{C}^l_{12}(x_{12}) \p_\lambda\langle\Phi_l(\bar x)\prod_{n\geq3}\Phi_n(x_n)\,.\rangle\right)
\ee
Writing $\p_{x_{1,2}}=\frac12\p_{\bar x}\pm\p_{x_{12}}$, 
the first term on the second line of (\ref{4.3d}) is $\propto\frac14\nabla^2_{\bar x}-\nabla^2_{12}$. Of these two pieces, the first modifies the coupling to $\nabla^2\Phi_l$, and therefore the conformal block, while the second is a contribution to
$\p_\lambda[{C}^l_{12}(x_{12})]$. The remaining terms, in the above limit,  contribute correctly to 
$\p_\lambda\langle\Phi_l(\bar x)\prod_{n\geq3}\Phi_n(x_n)\rangle$. Thus we have, to first order,
\be\label{Cdef}
\p_\lambda{C}^l_{mn}(x)=\log(\mu|x|)\nabla^2_x{C}^l_{mn}(x)\,,
\ee
thus simply generalizing (\ref{2ptx}) for the two-point function.

To go beyond this, first consider the deformed OPE in the cut-off bare theory
\be\label{bareope}
\Phi^\lambda_m(x_1)\cdot\Phi^\lambda_n(x_2)=\sum_l{C^\lambda}^l_{mn}(x_1-x_2;\ve)\Phi^\lambda_l(\bar x)\,,
\ee
where $\bar x=(x_1+x_2)/2$.
Now apply $\p_\lambda$ to both sides, using the Leibniz rule and (\ref{3.13}), and defining the string operator
$S^b(x)\equiv 2\pi\e^{ab}\e^{ij}\int_x^X dx'_jT^\lambda_{ai}(x'+\ve)$
\begin{eqnarray}
&&\p_\lambda\left(\Phi^\lambda_m(x_1)\cdot\Phi^\lambda_n(x_2)\right)
=\e^{ab}\left(S^b(x_1)\p_{x_1^b}+S^b(x_2)\p_{x_2^b}\right)\left(\Phi^\lambda_m(x_1)\cdot\Phi^\lambda_n(x_2)\right)\nonumber\\
&=&\ffrac12\e^{ab}\left((S^b(x_1)+S^b(x_2))(\p_{x_1^b}+\p_{x_2^b})+(S^b(x_1)-S^b(x_2))(\p_{x_1^b}-\p_{x_2^b})\right)\\
&&\times\sum_l{C^l_{mn}}^\lambda(x_1-x_2)\Phi^\lambda_l((x_1+x_2)/2)\\
&=&\e^{ab}S^b(x_1)\p_{x_1^b}(\Phi^\lambda_m(x_1))\cdot\Phi^\lambda_n(x_2)+\Phi^\lambda_m(x_1)\cdot \e^{ab}S^b(x_2)\p_{x_2^b}(\Phi^\lambda_n(x_2))\\
&=&\e^{ab}\left(S^b(x_1)\p_{x_1^b}+S^b(x_2)\p_{x_2^b}\right)\sum_l{C^l_{mn}}^\lambda(x_1-x_2)\Phi^\lambda_l((x_1+x_2)/2)\\
&=&\e^{ab}\sum_l\left((S^b(x_1)-S^b(x_2)){\p_bC^l_{mn}}^\lambda(x_1-x_2)\Phi^\lambda_l((x_1+x_2)/2)\right.
\label{4.4}\\
&&+[(1/2)(S^b(x_1)+S^b(x_2))-S^b((x_1+x_2)/2)]{C^l_{mn}}^\lambda(x_1-x_2)
\p_b\Phi^\lambda_l((x_1+x_2)/2))\label{4.5}\\
&&\left.+(S^b((x_1+x_2)/2){C^l_{mn}}^\lambda(x_1-x_2)\p_b\Phi^\lambda_l((x_1+x_2)/2))\right)\label{4.6}\,.
\end{eqnarray}
We recognize the last line (\ref{4.6}) as involving the deformed field $\Phi^\lambda_l$, while (\ref{4.4}, \ref{4.5}) describe the evolution of the OPE coefficients. (\ref{4.5}) involves the derivative field $\p_b\Phi^\lambda_l$,  so may be seen as a correction to the conformal block. (\ref{4.4}) thus gives the evolution of the coefficients:
\be
\p_\lambda {C^l}_{mn}^\lambda(x_1-x_2)=2\pi e^{ab}\int_{x_1}^{x_2}T_{ai}(x'+\ve)\e^{ij}dx'_j\,\p_b{C^l}_{mn}^\lambda(x_1-x_2)\,,
\ee
which, however, is still field valued, acting on $\Phi^\lambda_l((x_1+x_2)/2))$. It is also logarithmically divergent.
However, this is canceled in passing to the renormalized version in which $\Phi_n$ is replaced by
$\widehat\Phi_n$ and $C^l_{mn}$ by $\widehat C^l_{mn}$, leaving behind a factor of $\log(\mu|x_{12}|)$. The less singular terms in the OPE then contribute to descendent fields. Therefore (\ref{Cdef}) is exact for deformed primary fields, although the conformal blocks deform nontrivially. 

This result may also be derived from the RG, demanding that the bare OPE (\ref{bareope}) be independent of $\mu$. This leads to a Callan-Symanzik equation for the OPE coefficients, with a solution in $k$-space (see \ref{ck})
\be
{\widehat C}_{mn}^l(k,\lambda,\mu)=c^l_{mn}k^{\Delta_m+\Delta_n-\Delta_l}(k/\mu)^{2\lambda k^2}+\cdots\,,
\ee
in agreement with (\ref{Cdef}). Both of these exhibit unphysical short-distance behavior if $\lambda>0$.

\section{Green function method}\label{sec:GF}
We now describe an alternative method for decoupling the $T\overline T$ and similar deformations,  which leads to the same main equation (\ref{3.13}) but avoids the use of semi-local fields and gives a closer correspondence to the CFT calculation of Sec.~\ref{sec:CFT}. It also works in some other geometries. As before, we need to evaluate
$\int\det Td^2x$, regularized by point splitting, when inserted into a correlation function. We may write this as
\be\label{6.1}
2\pi\int\e^{ab}\e^{ij}\delta^{(2)}(x-x')T^\lambda_{ai}(x+\ve)T^\lambda_{bj}(x')d^2xd^2x'\,.
\ee
Introducing the Coulomb Green function satisfying $-\nabla^2_xG(x-x')=\delta^{(2)}(x-x')$, with suitable boundary conditions, this becomes
\be\label{6.2}
2\pi\int\e^{ab}\e^{ij}[\p_{x^k}\p_{x'_k}G(x-x'-\ve)]T^\lambda_{ai}(x)T^\lambda_{bj}(x')d^2xd^2x'\,.
\ee
 Now use
$\e^{ij}\p_{x'_k}=\e^{kj}\p_{x'_i}+\e^{ik}\p_{x'_j}$ ($\equiv \e^{jk}\p_{x_i}+\e^{ik}\p_{x'_j}$ when acting on $G(x-x')$), and integrate by parts to get two terms
\be
-2\pi\int\e^{ab}\e^{ik}[\p_{x^k}G(x-x'-\ve)]T^\lambda_{ai}(x)[\p_{x'_j}T^\lambda_{bj}(x')]d^2xd^2x'
\ee
\be
-2\pi\int\e^{ab}\e^{jk}[\p_{x^k}G(x-x'-\ve)][\p_{x_i}T^\lambda_{ai}(x)]T^\lambda_{bj}(x')d^2xd^2x'\,.
\ee
Under the interchange $(xai)\leftrightarrow(x'bj)$, these two terms are equal aside from $\ve\to-\ve$. 

We may now use the Ward identity (\ref{3.13d}) to evaluate $\p_{x'_j}T^\lambda_{bj}(x')$ acting on a general correlator $\langle\prod_n\Phi_n(x_n)\rangle$. There are terms where this acts on $T^\lambda_{ai}(x)$, which however vanish:
\be\label{4.5e}
\int\e^{ab}\e^{ik}[\p_kG(-\ve)]\p_bT_{ai}(x)d^2x+(\ve\to-\ve)=0\,.
\ee
 Acting on 
$\prod_p\Phi_p(x_p)$ it gives
\be\label{6.4e}
-2\pi\int d^2x\e^{ab}\e^{ik}\sum_n[\p_{x^k}G(x-x_n)]T^\lambda_{ai}(x+\ve)\,\p_{x_n^b}+(\ve\to-\ve)\,.
\ee
The connection with (\ref{3.13}) is now found by writing, in the case of the full plane, $\p_{x^k}G(x-x_n)=(x-x_n)_k/|x-x_n|^2$ and noting that, by conservation, the radial integral
\be
\int_0^\infty\e^{ik}((x-x_n)_k/|x-x_n|)T^\lambda_{ai}(x+\ve)d|x-x_n|
\ee
is in fact independent of the direction of $x-x_n$. Thus, apart from a factor of $2\pi$, we may fix a particular direction. 
(\ref{6.4e}) is then equivalent to (\ref{3.13}) if we take $S_n$ to lie along the radial direction from $x_n$ to $X=\infty$. 

However, (\ref{6.4e}) may be manipulated further by a second integration by parts, giving
\be\label{4.8f}
\p_\lambda\langle\prod_p\Phi_p(x_p)\rangle_\lambda=
2\pi\int d^2x\e^{ab}\e^{ik}\sum_nG_\ve(x-x_n)\p_k T^\lambda_{ai}(x+\ve)\,\p_{x_n^b}
\ee
\be
=2\pi\int d^2x\sum_nG_\ve(x-x_n)\p_{x^b}T^\lambda_{aa}(x+\ve)\,\p_{x_n^b}
-2\pi\int d^2x\sum_nG_\ve(x-x_n)\p_{x^a}T^\lambda_{ab}(x+\ve)\,\p_{x_n^b}\,,
\ee
where $G_{\ve}(x)=(1/4\pi)\log(|x|^2+|\ve|^2)$.
Using  the Ward identity  again in the second term, 
\begin{eqnarray}
\p_\lambda\langle\prod_p\Phi_p(x_p)\rangle_\lambda
&&=-\int d^2x\sum_{n}\frac{(x^b-x^b_n)\p_{x_n^b}}{|x-x_n|^2+\ve^2}\langle {T^\lambda}_a^a(x)\prod_p\Phi_p(x_p)\rangle_\lambda\nonumber\\
&&-\sum_{m,n}G_\ve(x_m-x_n)\,\p^{x_m^b}\p_{x_n^b}\langle \prod_p\Phi_p(x_p)\rangle_\lambda+(\ve\to-\ve)\,.
\label{4.10e}
\end{eqnarray}
The second term generalizes the CFT result (\ref{3.7}) to finite $\lambda$, and the first gives non-local corrections to it when the trace ${T^\lambda}_a^a\not=0$. However, from (\ref{3.13d}), it is non-singular as $\ve\to0$. 

\subsection{Other geometries}
For an infinitely long cylinder, parametrized by $-\infty<x^1<\infty$ and $0\leq x^2<R$ identified periodically, it is simpler to  decouple (\ref{6.1}) only in $(x^1,x'^1)$ by using the Green function 
$G=|x^1-x'^1|$. Integrating over $x'^1$ by parts from $x^1$ to $x^1+y^1$ now leads to
\be
\e^{ab}\e^{ij}\int T^\lambda_{ai}(x^1,x^2)T^\lambda_{bj}(x^1,x^2)dx^1dx^2
=\e^{ab}\e^{ij}\int T^\lambda_{ai}(x^1,x^2)T^\lambda_{bj}(x^1+y^1,x^2)dx^1dx^2
\ee
for all $y^1$, which is Zamolodchikov's identity, leading to the well-known Burgers equation for the energy levels. Note that if we apply this method to a correlation function $\langle \Phi(x_1)\Phi_(x_2)\rangle$ we need to evaluate the string expectation value
\be
\int_{x_1+\ve}^{x_2-\ve}\langle \Phi(x_1)T_{22}(x')\Phi(x_2)\rangle dx'^1\,,
\ee
which is non-trivial, since $T_{22}$ is not diagonal in the energy eigen-basis.

For a torus, the Coulomb Green function is not single-valued, so we already get a contribution for the partition function. The computation and result is similar to that in \cite{JC1} and we do not repeat it here.

\section{Interpretation as a field-valued diffeomorphism}\label{sec:diffeo}

We now 
return to the main result (\ref{3.13}) and reinterpret it as a field-dependent coordinate transformation. This point of view has been extensively discussed at the classical level, and extended to similar deformations in integrable models, by Conti, Negro and Tateo \cite{Conti1,Conti2}. 
\be
\p_\lambda\Phi^\lambda(x)
=2\pi\e_{ab}\e^{ij}\int_{x}^X dx'_j{T^\lambda}^a_i(x'+\ve)\p_{x^b}\Phi^\lambda(x)\,.
\ee
Because of the derivative $\p_{x^b}$ acting on $\Phi^\lambda(x)$
the first term may be interpreted formally as a change of coordinates 
\be\label{5.2}
\p_\lambda x_b=2\pi\e_{ab}\e^{ij}\int_{x}^X {T^\lambda}^a_i(x'+\ve)dx'_j\,.
\ee
Of course this is field-valued, and makes sense only inside correlation functions. Note also that we are here treating $\Phi^\lambda(x)$ as a bare field, to be inserted into the path integral with a UV cut-off, rather than as a renormalized field which would in general transform non-trivially under a diffeomorphism. Note however that
$\p_a\p_\lambda x_b=\p_b\p_\lambda x_a$, so there is no local rotation, and therefore it is immaterial whether $\Phi^\lambda(x)$ is a scalar or has higher rank under rotations. 

(\ref{5.2}) corresponds to a change in the (flat) metric
\be
\p_\lambda g_{ai}=\p_{x^a}(\p_\lambda x_i)+\p_{x^i}(\p_\lambda x_a)=4\pi\e_{ab}\e_{ij}{T^\lambda}^{bj}(x+\ve)\,,
\ee
which is just the saddle-point equation derived in \cite{JC1} when an infinitesimal $\det T$ deformation was decoupled by a gaussian random metric. 

However its is interesting to interpret (\ref{5.2}) equivalently in hamiltonian quantization as a state-dependent coordinate transformation. Labeling the coordinates now as $(x^0,x^1)$, quantizing along $x^0=0$ and running the integration along $x'^0=0, x'^1\geq x^1$ with $X=(0,+\infty)$, (\ref{5.2}) becomes
\begin{eqnarray}
\p_\lambda x^1&=&2\pi\int_{x^1+\ve}^\infty T^\lambda_{00}(0,x'^1)dx'^1\,,\\
\p_\lambda x^0&=&-2\pi\int_{x^1+\ve}^\infty T^\lambda_{10}(0,x'^1)dx'^1\,,
\end{eqnarray}
or, more symmetrically,
\begin{eqnarray}
\p_\lambda x^1&=&2\pi\left(\int_{x^1+\ve}^\infty-\int_{-\infty}^{x^1-\ve}\right) T^\lambda_{00}(0,x'^1)dx'^1=(E_>-E_<)\,,\\
\p_\lambda x^0&=&-2\pi\left(\int_{x^1+\ve}^\infty-\int_{-\infty}^{x^1-\ve}\right)  T^\lambda_{10}(0,x'^1)dx'^1=-(P_>-P_<)\,.
\end{eqnarray}
Thus, in a given state, the space coordinate is shifted according to the imbalance of total energy to its right and left, while the time coordinate is shifted according to the imbalance of momentum. In an asymptotic scattering state
with ordered energy-momenta $\{p^i_a\}$, a given particle will therefore suffer a time delay $\propto\sum_{b>a}p^1_b-\sum_{b<a}p^1_b$, leading to a total phase shift $\propto\lambda\sum_{a>b}\e_{ij}p_a^ip_b^j$, as already discovered for the $T\overline T$ deformation in several works, and equivalent to a gravitational  dressing\cite{Dub}. It is worth noting, however, that this applies also to non-relativistic systems. An interesting feature of this result is that the phase shift is exactly linear in the deformation parameter $\lambda$.

\section{Conclusions and further problems}
In this paper we have shown how the solvability of the $T\overline T$ deformation of a 2d quantum field theory (and of similar deformations) extends to correlation functions of local fields. Perhaps the most important conceptual result is that the deformation is a derivation on the algebra of such fields. This implies that the fusion rules of the undeformed UV CFT are preserved. More explicitly, we may consider a deformed correlation function to be equivalent to a correlation function of products of deformed fields with respect to the undeformed theory,
\be
\langle\prod_n\Phi_n(x_n)\rangle_\lambda=\langle\prod_n\Phi^\lambda_n(x_n)\rangle_{\lambda=0}\,,
\ee
where the deformation acts on the product according to the Leibniz rule
\be
\p_\lambda\big(\prod_n\Phi^\lambda_n\big)=\sum_m\big((\p_\lambda\Phi^\lambda_m)\prod_{n\not=m}\Phi^\lambda_n\big)\,.
\ee
However  the deformation of each field is non-local:
\be
\p_\lambda\Phi^\lambda(x)=S^b[x]\cdot\p_b\Phi^\lambda(x)\,,
\ee
which attaches a `string' $S^b[x]$ to $\Phi^\lambda(x)$ which is a line integral of $T^\lambda $. This effectively inserts an infinitesimally thin uniform strip running from $x+\ve$ to infinity (or equivalently to another field insertion), across which the original coordinate system is discontinuous. This induces conical singularities in the euclidean metric, indicating that the analysis of R\'enyi entropies is likely to be subtle. 

The stress tensor $T^\lambda $ itself obeys a similar evolution equation, with an extra term which ensures that it continues to be conserved and to satisfy its Ward identities.  One of the outstanding problems is understanding, in a deformed  CFT, the fate of the Virasoro algebra and why the spectral degeneracies on the cylinder dictated by its representations persist.  However a problem which affects all these considerations is how properly to define the stress tensor, beyond simply demanding that its OPE satisfy the translational and rotational Ward identities. Conventionally this is done by defining $T^{ij}$ as the response of the action to a general variation of the metric $g_{ij}$, but in this case this would require a generalization of the $\det T$ deformation to curved space, in such a way that it remains solvable. This appears difficult, since, for example, the covariant conservation law $\nabla_jT^{ij}=0$ no longer implies that $T^{ij}$ may be written as the curl of a potential as in flat space. Other approaches also appear to fail, except at large $c$ when factorization holds trivially.

Despite the semi-local nature of the deformed fields, the main result (\ref{3.13}) allows an analysis of the UV divergences in correlation functions as the $T(x)\overline T(x+\ve)$ point splitting regulator $\ve\to0$. There are \em universal \em logarithmic divergences $\propto \log|\ve|\nabla^2\Phi^\lambda$ which occur for all fields $\Phi^\lambda$ (including conserved currents like $T^\lambda$, although not in a way as to spoil their Ward identities). 

The simply nested nature of these divergences and their universality allow the deformed theory, albeit non-local, to be renormalized to all orders. The renormalized correlation functions obey a deformed OPE and also an RG equation. The solution (\ref{ck}) of this in $k$-space is however unusual, and it leads to different short-distance behaviors in real space for $\lambda>0$ and $\lambda<0$ (although in both cases the CFT short-distance power law behavior is suppressed.) 

Another interesting question is whether and, if so, how the deformed renormalized correlation functions described here are related, on mass shell, to the deformed $S$-matrix with CDD factors already discussed in the literature \cite{Dub2,Zam2,Tat1}.

A possibly more profitable line of investigation follows on the interpretation of the deformation as a field-, or state-dependent diffeomorphism of flat space, as originally studied in the classical case in \cite{Conti1}. Acting on asymptotic particle states, this gives a simple derivation of the appearance of CDD factors. More interestingly, this approach  applies to other similar deformations and also to non-Lorentz invariant theories.

Finally, although in Sec.~\ref{sec:GF} we introduced a different way of decoupling the $T\overline T$ term in the action, the method using Airy potentials used in the main body of this work appears to be more versatile and may be applied to give new results in open geometries and applied to entanglement properties of the deformed vacuum. It is hoped to describe these in a future paper.

\acknowledgments
 The author acknowledges useful conversations with V.~Bulchandani, A.~Gromov,  M.~Mezei, S.~Shenker, 
E.~Silverstein and R.~Tateo. 
This work was supported in part by the Simons Foundation through the Chern-Simons initiative at the University of California, Berkeley. 
A preliminary version was presented in April 2019 at the workshop on `TTbar and Other Solvable Deformations of Quantum Field Theories' at the Simons Center for Geometry and Physics, and the author gratefully acknowledges  support from the Center.

\end{document}